\documentclass[12pt]{iopart}

\usepackage{amsfonts}
\usepackage{amssymb}
\usepackage[dvips]{graphicx}

\usepackage{bbm} 
\usepackage{dcolumn}% Align table columns on decimal point
\usepackage{bm}% bold math
\usepackage{graphicx}%
\usepackage{epsfig}
\usepackage{tikz}
\usetikzlibrary{arrows,decorations.pathmorphing,backgrounds,positioning,fit,petri}
\begin{document}

%\draft
%\twocolumn

%\letter{}
\title{A quantum model of almost perfect energy transfer}

\author{Robert Alicki}

\address{
Institute of Theoretical Physics and Astrophysics, University
of Gda\'nsk,  Wita Stwosza 57, PL 80-952 Gda\'nsk, Poland}
\ead{fizra@univ.gda.pl}

\author{Filippo Giraldi}
\address{Quantum Research Group,
School of Physics and National Institute for Theoretical Physics,
University of KwaZulu-Natal, Durban 4001, South Africa}
\ead{filgi@libero.it}

\begin{abstract}
The Wigner-Weisskopf-type model describing the energy transfer between two centers mediated by a continuum of energy levels is studied. This work is motivated by the recent interest in transport phenomena at nanoscale in biology and quantum engineering. The analytical estimation for the energy transfer efficiency is derived in the weak coupling regime and the conditions for the almost perfect transfer are discussed. The embedding of the standard tight-binding model into Wigner-Weisskopf one which includes the environmental noise is presented.
\end{abstract}

\pacs{03.65 Yz , 03.67.-a , 05.60 Gg}

%\submitto{\JPA}

\date{\today}
% \date{July 26, 2003}
\maketitle
\section{Introduction}
The almost perfect energy transfer (APET) in biologically relevant molecular systems attracts attention of physicists in the recent years \cite{May}.
Similar phenomena should be observed also in the engineered systems relevant for information processing like arrays of quantum dots or Josephson junctions \cite{Kea}. Most of the theoretical analysis is performed using tight-binding Hamiltonian for  an interacting N-body system in the single excitation regime. The influence of environment is usually modeled by Markovian master equations with phenomenological decay parameters. The numerical computations show the interplay between quantum propagation, quantum localization, decoherence and dissipation \cite{Ple,Reb,Wha}. The aim of this paper is to present a simple, essentially exactly solvable model which describes a generic class of such phenomena and allows to derive analytical bounds on the efficiency of energy transfer processes. The motivation for the applied formalism comes from the particular model of two
identical atoms placed in the focuses of two parabolic mirrors separated by a distance $\ell$, see Fig.1. We assume that the dipole moments of both atoms are parallel to the symmetry axis of the system.
\begin{figure}[h]
\centering
\begin{tikzpicture}
[place/.style={circle,draw=black!50,fill=black!20,thick,inner sep=0pt,minimum size=1mm},
emp/.style={circle,draw=white!50,fill=white!20,thick,inner sep=0pt,minimum size=0.1mm},
bend angle=20,
pre/.style={<-,shorten <=1pt,>=stealth,semithick},
post/.style={->,shorten >=1pt,>=stealth,semithick}]
%\coordinate [label=left:$A$] (A) at (-1,0);
%\draw (A);
\node[place]  (A1)  at (-1,0.5) {};
\node[place]  (A2)  at (7,0.5) {};
\node[emp]  (A3)  at (-1.5,0.5) {}
	edge [<-, , thick] node[auto] {} (A1);
\node[emp]  (A4)  at (-0.5,0.5) {}
	edge [<-, , thick] node[auto] {} (A1);
\node[emp]  (A5)  at (7.5,0.5) {}
	edge [<-, , thick] node[auto] {} (A2);
\node[emp]  (A6)  at (6.5,0.5) {}
	edge [<-, , thick] node[auto] {} (A2);
\node[emp]  (A7)  at (-1.5,1.25) {}
	edge [<-, , thick, dotted] node[auto] {} (A1);
\node[emp]  (A8)  at (-1.5,-0.25) {}
	edge [<-, , thick, dotted] node[auto] {} (A1);
\node[emp]  (A9)  at (7.5,1.25) {}
	edge [->, , thick, dotted] node[auto] {} (A2);
\node[emp]  (A10)  at (7.5,-0.25) {}
	edge [->, , thick, dotted] node[auto] {} (A2);
\node[emp]  (A11)  at (-0.5,1.75) {}
	edge [<-, , thick, dotted] node[auto] {} (A1);	
\node[emp]  (A12)  at (-0.5,-0.75) {}
	edge [<-, , thick, dotted] node[auto] {} (A1);
\node[emp]  (A13)  at (6.5,1.75) {}
	edge [->, , thick, dotted] node[auto] {} (A2);
\node[emp]  (A14)  at (6.5,-0.75) {}
	edge [->, , thick, dotted] node[auto] {} (A2);	
\draw [->,decorate, decoration={snake,amplitude=.4mm,segment length=2mm,post length=1mm}] (-1.3,1.25) -- (7.3,1.25);
\draw [->,decorate, decoration={snake,amplitude=.4mm,segment length=2mm,post length=1mm}] (-1.3,-0.25) -- (7.3,-0.25);
\draw [->,decorate, decoration={snake,amplitude=.4mm,segment length=2mm,post length=1mm}] (-0.4,1.75) -- (6.4,1.75);
\draw [->,decorate, decoration={snake,amplitude=.4mm,segment length=2mm,post length=1mm}] (-0.4,-0.75) -- (6.4,-0.75);											
\draw[rotate=90] (-1,0) parabola[bend pos=0.5] bend +(0,2) +(3,0);
\draw[rotate=-90] (-2,6) parabola[bend pos=0.5] bend +(0,2) +(3,0);
%\draw (-1,0) parabola[parabola height=2cm]
\end{tikzpicture}

\caption{The APET for two atoms places in the focuses of two parabolic mirrors. Radiating dipoles are parallel to the axis. }
\label{Mirrors}
\end{figure}
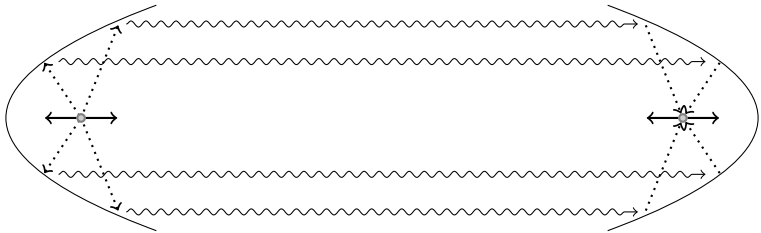
If initially the first atom is excited and the second is in the ground state we expect that roughly after  time
$t \simeq \ell/c + \mathcal{O}(\tau)$, where $\tau$ is a mean life-time of the excited state, the energy quantum carried by a photon is almost perfectly transfered to the second atom. The simple description of such a system can be given in terms of the Wigner-Weisskopf (W-W) Hamiltonian (compare the case of a single mirror and a single atom \cite{Sto}).
The conditions which allows for APET are encoded entirely in the spectrum of the Hamiltonian which is determined by the details of the model. Treating W-W model as a generic one of a large class of energy transfer phenomena between two localized centers one can try to find a general bound for the efficiency
of such process and determine the sufficient conditions for APET.
\section{Wigner-Weisskopf model}
In order to introduce a 2-state W-W model we consider a quantum system of  two 2-level atoms interacting with an electromagnetic field which can be described by the standard Hamiltonian in the rotating wave approximation
\begin{eqnarray}
H &=& \frac{\omega_1}{2}\sigma^z_1 + \frac{\omega_2}{2}\sigma^z_2 + \int d^3\mathbf{k}\, \omega(\mathbf{k})\,a^{\dagger}(\mathbf{k}) a(\mathbf{k})
\nonumber\\
&+& \sigma_1^{+}a(g_1) + \sigma_1^{-}a^{\dagger}(g_1) +\sigma_2^{+}a(g_2) + \sigma_2^{-}a^{\dagger}(g_2) .
\label{ham_2}
\end{eqnarray}
The operators $a(g)$ and  $a^{\dagger}(g) $ are smeared annihilation and creation operators for the electromagnetic field given by $a^{\dagger}(g)=\int d^3\mathbf{k}\, g(\mathbf{k})a^{\dagger}(\mathbf{k})$, $g_{j}(\mathbf{k}), j=1,2$ are suitable formfactors, and the quantum fields satisfy $[a(\mathbf{k}),a^{\dagger}(\mathbf{k'})]= \delta(\mathbf{k}-\mathbf{k'})$ (we omit for simplicity the polarization degrees of freedom). The standard $2\times 2 $ Pauli matrices $\sigma_{1,2}^{\pm,z}$ refer to 2-level atoms. In the case of identical atoms with relative position described by the vector $\mathbf{r}$ the formfactors differ by the relative phase $L(\mathbf{k})=\mathbf{k\cdot r}$
\begin{equation}
g_2(\mathbf{k}) =  e^{-iL(\mathbf{k})}g_1(\mathbf{k}).
\label{relshift}
\end{equation}
The crucial property of the Hamiltonian (\ref{ham_2}) is the commutation with the excitation number operator defined as
\begin{equation}
N_{ex} = \int d^3\mathbf{k}\, a^{\dagger}(\mathbf{k}) a(\mathbf{k})+ (\sigma^z_1 +1/2) + (\sigma^z_2 +1/2).
\label{exN}
\end{equation}
The total Hilbert space of the system possesses a tensor product structure
\begin{equation}
\mathcal{H}_{tot} = {\mathbb C}^2\otimes \mathcal{F}_{ph}\otimes {\mathbb C}^2
\label{Htotal}
\end{equation}
with the bosonic Fock space $\mathcal{F}_{ph}$ describing an electromagnetic field.
As a consequence of the commutation $[H , N_{ex}] =0$ the subspace corresponding to the eigenvalue $1$ of $N_{ex}$, which can be called single exciton Hilbert space is invariant under the evolution. It possesses a direct orthogonal sum structure, i.e. is spanned by the following vectors
\begin{eqnarray}
|1\rangle &\equiv& |\uparrow;1\rangle \otimes|\mathrm{vac}\rangle\otimes |\downarrow;2\rangle ,
\nonumber \\
|2\rangle &\equiv &|\downarrow;1\rangle \otimes |\mathrm{vac}\rangle\otimes |\uparrow;2\rangle ,
\nonumber \\
|f\rangle &\equiv& |\downarrow;1\rangle \otimes a^{\dagger}(f)|\mathrm{vac}\rangle\otimes |\downarrow;2\rangle   \ \mathrm{for\ any\ wave-packet}\ f.
\label{exciton}
\end{eqnarray}
Here, $|\uparrow;j\rangle$ and $|\downarrow;j\rangle$ denote excited  and ground state of the $j$-th atom, respectively, and $|\mathrm{vac}\rangle$ is the vacuum state of an electromagnetic field. This mathematical construction, which has been used frequently in quantum optics \cite{Aga}, will be called the 2-state W-W model.
\par
The 2-state W-W model is suitable for the following generic physical situation. We restrict  ourselves to physical systems which can be described in terms of the single-exciton Hilbert space $\mathcal{H}_{ex}$.
In this Hilbert space we choose two orthogonal vectors $|1\rangle, |2\rangle$ which are determined by the initial state preparation and the measurement procedures. Namely, the initial state ("donor") of the system at time $t_0 =0$ is denoted by $|1\rangle$,
while the measurement at time $t>0$ after preparation is a von Neumann projection on the  state $|2\rangle$ ("acceptor").
\par
The Hilbert space of the model system is decomposed into a direct sum (compare with (\ref{exciton}))
\begin{equation}
\mathcal{H}_{ex} = {\mathbb C}\oplus \mathbf{L}^2(\Omega)\oplus {\mathbb C}
\label{Hspace}
\end{equation}
where the one-dimensional subspaces ${\mathbb C}$ are generated by the states $|1\rangle ,|2\rangle$ and the Hilbert space $\mathbf{L}^2(\Omega)$ is their orthogonal supplement. For the convenience we represent this Hilbert space as the space of wave packets $f(k), g(k), k\in\Omega $ denoted by $|f\rangle , |g\rangle$ with the scalar product
\begin{equation}
\langle f|g\rangle = \int_{\Omega}\overline{f(k)}g(k)\, dk\ .
\label{scalar}
\end{equation}
One should stress that the notation of above is used for convenience only. One can always replace the continuous variable $k$ by a joint set of continuous variables $\omega$ and discrete quantum numbers $m$ such that $|k\rangle$ is replaced by
$|\omega , m\rangle$  and
\begin{equation}
\langle \omega, m|\omega',m'\rangle = \delta(\omega-\omega')\delta_{mm'}\ ,\ \mathrm{and} \   \int_{\Omega} dk \mapsto \sum_m \int d\omega .
\label{para}
\end{equation}
The 2-state W-W Hamiltonian, which can be seen as the restriction of (\ref{ham_2}) to the single-exciton space, reads
\begin{eqnarray}
H &=& H_0 + V,
\nonumber \\
H_0&=& \omega_1 |1\rangle\langle 1| + \omega_2 |2\rangle\langle 2| +\int_{\Omega} dk\,\omega(k)|k\rangle\langle k|,
\nonumber \\
V &=& (|1\rangle\langle g_1|+ |g_1\rangle\langle 1|)+ (|2\rangle\langle g_2|+ |g_2\rangle\langle 2|)
\label{WWham}
\end{eqnarray}
where $g_{1,2}\in \mathbf{L}^2(\Omega)$. All details of the model are hidden in the form of formfactors $g_{1,2}$ and the spectral resolution of the part of the Hamiltonian denoted here by $\int_{\Omega} dk\,\omega(k)|k\rangle\langle k|$. The only assumptions used in the following calculations are the continuity of the spectrum $\{\omega(k)\}$ and the orthogonality  $\langle g_1|g_2\rangle =0$. The later condition means that there is no direct cross-talking between the preparation and measurement procedures and simplifies the calculations.

\section{Efficiency of energy transfer}
The system begins its evolution in the excited state $|1\rangle $ and after time $t$ can be found in the state $|2\rangle$ with the probability
\begin{equation}
\mathcal{P}_{12}(t) = |\mathcal{A}_{12}(t) |^2 \ ,\ \mathcal{A}_{12} =\langle 2|e^{-iHt} |1\rangle.
\label{Pro}
\end{equation}
The APET holds if $\omega_1\simeq \omega_2$ and for a certain time $t_0$ the transfer probability $\mathcal{P}_{12}(t_0)\simeq 1$. To find the sufficient conditions for APET in terms of the Hamiltonian (\ref{WWham}) notice that the (complex) probability amplitude $\mathcal{A}_{12}(t)$  is a scalar product of two states, $e^{-iHt/2}|1\rangle$ and $e^{iHt/2}|2\rangle$  which evolve forward and backward in time, respectively. For $t \gg\tau$ where $\tau$ is the life-time of the excited states one expects that the following approximation holds
\begin{equation}
e^{-iHt/2}|1\rangle \simeq e^{-iH_0 t/2}|f_1\rangle\ ,\  e^{iHt/2}|2\rangle \simeq e^{iH_0 t/2}|f_2\rangle
\label{est1}
\end{equation}
where $f_1(k)$, $f_2(k)$ are certain wave packets from the Hilbert space component $\mathbf{L}^2(\Omega)$ describing intermediate excitonic states. The estimation (\ref{est1}) is equivalent to the existence of wave operators defined as
\begin{equation}
W_{+} = \lim_{t\to+\infty} e^{iH_0t} e^{-iHt}\ ,\ W_{-} = \lim_{t\to+\infty} e^{-iH_0t} e^{iHt}
\label{Wop}
\end{equation}
which holds under  mild conditions on the formfactors $g_{1,2}$ and exciton's dispersion relation $\omega(k)$, at least in the weak sense of convergence of matrix elements. Combining (\ref{est1}) with (\ref{Wop}) one obtains
\begin{equation}
f_1(k) = \langle k|W_{+}|1\rangle \ ,\ f_2(k) = \langle k|W_{-}|2\rangle
\label{f_12}
\end{equation}
and finally
\begin{equation}
\mathcal{A}_{12}(t)= \int_{k\in\Omega} dk\, e^{-i\omega(k)t} \langle k|W_{+}|1\rangle \overline{ \langle k|W_{-}|2\rangle}.
\label{A_12}
\end{equation}
The advantage of the Wigner-Weisskopf model is the fact that the matrix elements $\langle 1|W_- |k\rangle$ and $\langle 2|W_+ |k\rangle$ can be exactly computed using Laplace transforms (see the next Section).
\section{The computation of matrix elements}
\par
The basic tools used for the computation of the probability amplitude  (\ref{A_12}) are:\\
the  identity
\begin{eqnarray}
e^{-iHt} &=& e^{-iH_0 t} -i\int_0^t ds\, \big[ e^{-i\omega_1(t-s)}|1\rangle\langle g_1| + |g_1(t-s)\rangle\langle 1|
\nonumber \\
         & +& e^{-i\omega_2(t-s)}|2\rangle\langle g_2| + |g_2 (t-s)\rangle\langle 2|\bigr] e^{-iHs}
\label{ide}
\end{eqnarray}
where $|g_j (t)\rangle \equiv e^{-iH_0 t }|g_j\rangle$, and the Laplace transform
\begin{equation}
\tilde{f}(z) = \int_0^{\infty} e^{-zt}f(t)\, dt .
\label{laplace}
\end{equation}
Introducing the  notation  ($j, j'=1,2$):
\begin{eqnarray}
S_{jj'}(t) &=& \langle j| e^{-iH t}|j'\rangle
\nonumber \\
F_{jj'}(t) &=& \langle g_j| e^{-iH t}|j'\rangle
\nonumber \\
G_{jj'}(t) &=& \langle g_j| e^{-iH_0 t}|g_{j'}\rangle .
\label{def}
\end{eqnarray}
and using the identity (\ref{ide}), the definition of the wave operators (\ref{Wop}),  the notation (\ref{def}) and the Laplace transform we have
\begin{eqnarray}
f_1(k)=\langle k|W_+ |1\rangle &=& -i\left[g_1(k) \tilde{S}_{11}\left(-i \omega(k)\right)+g_2(k) \tilde{S}_{21}\left(-i \omega(k)\right)\right]
\nonumber \\
f_2(k)=\langle k|W_- |2\rangle &=&  i \left[g_1(k)\overline{ \tilde{S}_{21}\left(-i\omega(k)\right)}+g_2(k) \overline{\tilde{S}_{22}\left(-i \omega(k)\right)}\right].
\label{f2}
\end{eqnarray}
Combining the identity (\ref{ide}) with the definitions (\ref{def}) we obtain a series of equations
\begin{eqnarray}
S_{11}(t) &=& e^{-i\omega_1 t} -i\int_0^t e^{-i\omega_1 (t-s)} F_{11}(s)\, ds,
\nonumber \\
S_{22}(t) &=& e^{-i\omega_2 t} -i\int_0^t e^{-i\omega_2 (t-s)} F_{22}(s)\, ds,
\nonumber \\
S_{12}(t) &=&  -i\int_0^t e^{-i\omega_1 (t-s)} F_{12}(s)\, ds,
\nonumber \\
F_{11}(t) &=&  -i\int_0^t  G_{11}(t-s)S_{11}(s)\, ds,
\nonumber \\
F_{22}(t) &=&  -i\int_0^t  G_{22}(t-s)S_{22}(s)\, ds,
\nonumber \\
F_{12}(t) &=&  -i\int_0^t  \bigl[ G_{11}(t-s)S_{12}(s)+ G_{12}(t-s)S_{22}(s)\bigr] \, ds,
\label{eqs}
\end{eqnarray}
which can be  converted into equations for Laplace transforms
\begin{eqnarray}
\tilde{S}_{11}(z) &=& \frac{1}{z + i\omega_1}\bigl[1-i\tilde{F}_{11}(z)\bigr],
\nonumber \\
\tilde{S}_{22}(z) &=& \frac{1}{z + i\omega_2}\bigl[1-i\tilde{F}_{22}(z)\bigr],
\nonumber \\
\tilde{S}_{12}(z) &=& \frac{-i}{z + i\omega_1}\tilde{F}_{12}(z),
\nonumber \\
\tilde{F}_{11}(z) &=&  -i \tilde{G}_{11}(z)\tilde{S}_{11}(z),
\nonumber \\
\tilde{F}_{22}(z) &=&  -i \tilde{G}_{22}(z)\tilde{S}_{22}(z),
\nonumber \\
\tilde{F}_{12}(z) &=&  -i \bigl[\tilde{G}_{11}(z)\tilde{S}_{12}(z) \tilde{G}_{12}(z)\tilde{S}_{22}(z)\bigr] .
\label{eqsL}
\end{eqnarray}
The system of equations (\ref{eqsL}) can be  solved yielding the following basic formulas
\begin{eqnarray}
\tilde{S}_{11}(z) &=& \bigl[{z + i\omega_1+\tilde{G}_{11}(z)}\bigr]^{-1},
\nonumber \\
\tilde{S}_{22}(z) &=& \bigl[{z + i\omega_2 +\tilde{G}_{11}(z)}\bigr]^{-1},
\nonumber \\
\tilde{S}_{12}(z) &=& -\tilde{G}_{12}(z)\bigl[{z + i\omega_1+\tilde{G}_{11}(z)}\bigr]^{-1}\bigl[{z + i\omega_2 +\tilde{G}_{11}(z)}\bigr]^{-1}.
\label{solL}
\end{eqnarray}
Combining (\ref{solL} ) with (\ref{f2}) and  (\ref{A_12}) one can obtain the final formula for the transition amplitude which can be used for numerical calculations.
\section{Markovian approximation}
In order to continue analytical analysis of the problem we consider the case of weak coupling or Markovian approximation. Treating $V$ given by (\ref{WWham}) as a small perturbation one can omit in (\ref{f2}) the terms proportional to $\tilde{S}_{21}$. The further approximation concerns Laplace transforms $\tilde{G}_{jj}(z)$. The real part of $\tilde{G}_{jj}(-\omega_j)$ is a standard \emph{Fermi Golden Rule} approximation for the decay rate of the state
$|j\rangle$ ($j=1,2$) given by
\begin{equation}
\gamma_j = \gamma_j(\omega_j)\ ,\ \gamma_j(\omega)= \pi\int_{\Omega}dk\, |g_j(k)|^2 \delta(\omega (k)-\omega),
\label{dec}
\end{equation}
while the imaginary part is a \emph{radiative correction} to the bare frequencies $\omega_j$. In the following we denote by the same symbol $\omega_j$ the physical, renormalized frequencies. Summarizing, the Markovian approximation
means that
\begin{equation}
\tilde{S}_{21}(-i\omega(k))\simeq 0\ ,\ \tilde{S}_{jj}(-i\omega(k)) \simeq\bigl[i(\omega_j -\omega(k))+\gamma_j\bigr]^{-1}\ ,\ \gamma_j \ll \omega_j .
\label{mar}
\end{equation}
Therefore using Eqs. \ref{f2}) and (\ref{mar}) one obtains
\begin{equation}
f_1(k) = \frac{g_1(k)}{\omega(k) -\omega_1 +i\gamma_1},  \hspace{2em}\ f_2(k) = \frac{g_2(k)}{\omega(k) -\omega_2 -i\gamma_2}
\label{f_12mar}
\end{equation}
and
\begin{equation}
\mathcal{A}_{12}(t)= \int_{k\in\Omega} dk\, e^{-i\omega(k)t} \frac{g_1(k)\overline{g_2(k)}}
{(\omega(k) -\omega_1 +i\gamma_1)(\omega(k) -\omega_2 +i\gamma_2)}
\label{A_12mar}
\end{equation}
One can estimate the upper bound for $\mathcal{P}_{12}(t)$ using (\ref{A_12mar}) to get
\begin{equation}
\mathcal{P}_{12}(t)=|\mathcal{A}_{12}(t)|^2 \leq \|f_2\|^2\|f_1\|^2 .
\label{bound}
\end{equation}
The equality $\int_{-\infty}^{\infty}\gamma[x^2 +\gamma^2]^{-1}dx =\pi$ implies for example
\begin{eqnarray}
\|f_1\|^2 = \int_{\Omega}|f_1(k)|^2\,dk &=& \int_{\Omega}dk\, \frac{|g_1(k)|^2}{(\omega_1 -\omega(k))^2 +\gamma_1^2}
\nonumber\\
&\simeq& \frac{1}{\pi}\int_0^{\infty} \frac{\gamma_1}{(\omega_1 -\omega)^2 +\gamma_1^2}\simeq 1- \frac{\gamma_1}{\pi\omega_1}
\label{norm}
\end{eqnarray}
what finally gives the bound (compare with \cite{Ali})
\begin{equation}
\mathcal{P}_{12}(t)\leq \Bigl(1- \frac{\gamma_1}{\pi\omega_1}\Bigr)\Bigl(1- \frac{\gamma_2}{\pi\omega_2}\Bigr)< 1.
\label{bound1}
\end{equation}
In the weak coupling regime the upper bound is close to one.
\section{Conditions for APET}
In order to approach the bound (\ref{bound1}) and achieve  APET  certain  matching conditions implied directly by the formula (\ref{A_12mar}) must be satisfied. To present them in a transparent form we introduce certain additional assumptions
in a convenient parametrization. There are satisfied by the two-mirror system discussed in the Introduction.
\par
Assume that the intermediate Hilbert space $\mathbf{L}^2(\Omega)$ is spanned by the basis
$|\omega , m\rangle$ and the corresponding part of the Hamiltonian reads
\begin{equation}
H_{1}= \sum_m \int d\omega \,\omega\, |\omega , m\rangle \langle \omega , m| .
\label{hamred}
\end{equation}
The formfactors possess the following structure
\begin{equation}
|g_1\rangle = \int d\omega \, g(\omega) |\omega , m_0\rangle \ ,\ |g_2\rangle = \int d\omega \, e^{-iL(\omega)}g(\omega) |\omega , m_0\rangle .
\label{fred}
\end{equation}
where $g(\omega)$ is a "flat" slowly varying function and $L(\omega)$ accounts for the spatial separation of  donor and acceptor similarly to (\ref{relshift}). They satisfy approximative orthogonality condition
\begin{equation}
\int d\omega \, e^{iL(\omega)}|g(\omega)|^2 \simeq 0 .
\label{sep}
\end{equation}
Now the mechanism leading to APET can be illustrated. We have to put the resonance condition
\begin{equation}
\omega_1 = \omega_2 \equiv \omega_0,
\label{omeg}
\end{equation}
which by (\ref{dec}) implies also the equality of decay rates
\begin{equation}
\gamma_1 = \gamma_2 \equiv \gamma.
\label{gam1}
\end{equation}
Then we have the approximative expression
\begin{equation}
\mathcal{A}_{12}(t)\simeq \frac{1}{\pi}\int d\omega\, \frac{\gamma}{(\omega - \omega_0)^2 +\gamma^2}\Bigl(\frac {\omega - \omega_0 -i\gamma}{\omega - \omega_0 +i\gamma}\Bigr)e^{i(L(\omega) - \omega t)} .
\label{A_12app}
\end{equation}
Further approximations valid for $|\omega -\omega_0|\leq \gamma$
\begin{equation}
\Bigl(\frac {\omega - \omega_0 -i\gamma}{\omega - \omega_0 +i\gamma}\Bigr)\simeq -e^{i2(\omega-\omega_0)/\gamma}\ ,\ L(\omega)\simeq L(\omega_0) +L'(\omega_0)(\omega-\omega_0)
\label{approx}
\end{equation}
lead to
\begin{equation}
|\mathcal{A}_{12}(t)|\simeq \frac{1}{\pi}|\int d\omega\, \frac{\gamma}{(\omega - \omega_0)^2 +\gamma^2} \exp\bigl\{i\bigl[L'(\omega_0)+ 2\tau - t\bigr](\omega-\omega_0)\bigr\}|
\label{approx1}
\end{equation}
where $\tau = 1/\gamma$. The transition probability is maximal and close to one if the oscillating factor in the integral of above vanishes what happens for the time
\begin{equation}
t_0 = L'(\omega_0)+ 2\tau
\label{arrive}
\end{equation}
which can be called optimal transfer time.
\section{Transport in quantum networks}
The standard tight-binding model of  energy transport in quantum networks which is applicable both to molecular systems and engineered ones consist of $N$ 2-level "atoms" described by the Hilbert space ${\mathbb C}^{2^N}$ and the Hamiltonian written in terms of Pauli matrices
\begin{equation}
H_{N}= \sum_{k=1}^N \omega_k\, \sigma^z_k + \sum_{k<l=2}^N \bigl(h_{kl}\,\sigma^+_k \sigma^-_l + h.c.\bigr)
\label{t_bHam}
\end{equation}
where $\{\omega_k \}$ are energies of the sites and  $\{h_{kl}; k < l\}$ are hopping amplitudes. Similarly to the example in the Section 2 the exciton number operator 
\begin{equation}
N_{ex} = \sum_{k=1}^N(\sigma^z_k +1/2) 
\label{exN1}
\end{equation}
commutes with $H_N$.
\par
Again a single exciton $N$-dimensional Hilbert space ${\mathbb C}^{N}$ is invariant with respect to the dynamics
and the corresponding restriction of the Hamiltonian $H_N$ is given by the $N\times N$
hermitian matrix $H =[h_{kl}]$ with the diagonal elements $h_{kk}= \omega_k$.  We attribute to the donor site an index "1" and to the acceptor site index "2". We assume that the direct hopping $1\leftrightarrow 2$ is negligible, i.e. $h_{12}=h_{21}=0$ and the two vectors $|g_1\rangle = [0,0,h_{31},h_{41},...,h_{N1}]$ and  $|g_2\rangle = [0,0,h_{32},h_{42},...,h_{N2}]$ are orthogonal. Under these conditions the Hamiltonian can be recast into the discrete version of  the W-W Hamiltonian
\begin{equation}
H = \omega_1 |e_1\rangle\langle e_1| + \omega_2 |e_2\rangle\langle e_2| + \bigl(|e_1\rangle\langle g_1| + |e_2\rangle\langle g_2|+\mathrm{h.c.}\bigr) + H_1 .
\label{discreteWW}
\end{equation}
Here $|e_1\rangle =[1,0,...,0]$, $|e_2\rangle =[0,1,...,0]$ and $H_1$ is a submatrix of $H$ with indices $k,l = 3,4,...,N$. $H_1$ can be written in its spectral decomposition form
\begin{equation}
H_1 = \sum_{\alpha =3}^{N}\epsilon_{\alpha} |\alpha\rangle\langle\alpha |
\label{specH_1}
\end{equation}
to get a full analogy with the continuous W-W Hamiltonian (\ref{WWham}). In the next step one should embed the discrete model into a continuous one. Physically, it means that the interaction with other (environmental) degrees of freedom is taken into account. This interaction transforms the eigenstates $|\alpha\rangle$ into resonances with finite spectral widths. The Hilbert space spanned by $\{|\alpha\rangle\}$ is replaced by $L^2({\mathbb R})$ and the Hamiltonian (\ref{specH_1}) by the multiplication operator by $\omega$ which can be formally written as $\int d\omega\, \omega\,|\omega\rangle\langle\omega|$.
As a consequence the vectors $\{|g_{j}\rangle ; j=1,2\}$ are replaced by the  wave functions $g_{j}(\omega)$ which provide  continuous envelopes for the discrete values $\{g_{j}(\epsilon_{\alpha}) =\mathcal{N}_j\langle\alpha|g_{j}\rangle , \alpha = 3,4,...,N\}$. The constants $\mathcal{N}_j$ are chosen to satisfy normalization condition
\begin{equation}
\langle g_{j}|g_{j}\rangle = \int|g_{j}(\omega)|^2 d\omega\ ,\ j=1,2 .
\label{norm1}
\end{equation}
The construction of a continuous envelope for absolute values
$|\langle\alpha|g_{j}\rangle |$ is simple as one can use e.g. Lorentz profiles to smear the eigenvalues $\{\epsilon_{\alpha}\}$.
However, the envelope of the relative phases denoted by $L(\omega)$ is more tricky and  depends on the geometry of the system (compare with (\ref{relshift})).
\par
Using the results of the Section 6 we can discuss the conditions which should be satisfied to achieve APET, at least in the weak coupling regime. There are two types of such conditions,  fine-tuning  and generic ones.\\
The fine-tuning conditions are :\\
a) the resonance condition (\ref{omeg}),\\
b) the equality (\ref{gam1}) which means that the formfactor functions $g_{j}(\omega), j=1,2$ must cross at the resonance energy $\omega_0$.\\
The generic conditions are the following:\\
c) the optimal transfer time (\ref{arrive}) should be short enough what implies that the relative phase function $L(\omega)$ must be smooth enough around the value $\omega_0$,\\
d) the widths of the resonances  should be larger than the nearest neighbor energy spacing for $\{\epsilon_{\alpha}\}$ in order to produce flat envelopes $g_{j}(\omega)$ what means that the interaction with an environment should be strong enough.\\
In the case of engineered networks the fine-tuning conditions imply a careful design while for biologically relevant systems one can imagine that the natural selection mechanism plays a crucial role.

\section{Concluding remarks}
The 2-state Wigner-Weisskopf model provides a versatile mathematical tool to study,  within the single exciton approach, the transport properties in complex molecular systems or engineered networks relevant for quantum information processing, both in the weak coupling (Markovian) and strong coupling (non-Markovian) regimes. The presented
version of this model can be modified to include exciton decay and  presence of a sink by adding imaginary decay rates $i\Gamma(k)$ , $i\Gamma_2$ to energies $\omega(k)$ and $\omega_2$, respectively. The W-W Hamiltonian (\ref{WWham}) possesses also entirely classical interpretation as a Hamiltonian of a certain quadratic system (two harmonic oscillators coupled to a continuum of harmonic ones). Therefore, the physical mechanisms leading to APET are coherent wave-like ones and not specifically quantum which could be attributed to any nontrivial manifestation of entanglement.

\textbf{ Acknowledgments} The authors thank Ilya Sinayskiy for the assistance. R.A. is supported by the Polish Ministry of Science and Higher Education, grant PB/2082/B/H03/2010/38.

\end{document}